\documentclass[fleqn,usenatbib]{mnras}
\usepackage[T1]{fontenc}
\usepackage{ae,aecompl}
\usepackage{graphicx}
\usepackage{amsmath}
\usepackage{amssymb}
\usepackage{color}

\def\apj{{\rm ApJ}}                 
\def\apjl{{\rm ApJ}}                
\def\apjs{{\rm ApJS}}               
\def\ao{{\rm Appl.~Opt.}}           
\def\aap{{\rm A\&A}}                
\def\mnras{{\rm MNRAS}}             
\def\prd{{\rm Phys.~Rev.~D}}        

\def\mnras{Mon.\ Not.\ R.\ Astron.\ Soc.\ }
\newcommand{\avg}[1]{\ensuremath{\langle #1 \rangle}}
\newcommand{\bma}{\begin{math}}
\newcommand{\ema}{\end{math}}
\newcommand{\beqa}{\begin{eqnarray}}
\newcommand{\eeqa}{\end{eqnarray}}
\newcommand{\bc}{\begin{center}}
\newcommand{\ec}{\end{center}} 
\newcommand{\bit}{\begin{itemize}}
\newcommand{\eit}{\end{itemize}}

\newcommand{\beq}{\begin{equation}}
\newcommand{\eeq}{\end{equation}}
\newcommand{\bea}{\begin{eqnarray}}
\newcommand{\eea}{\end{eqnarray}}
\definecolor{darkgreen}{cmyk}{0.85,0.2,1.00,0.2} 
\definecolor{darkblue}{cmyk}{1,1,0,0} 
\definecolor{purple}{cmyk}{0.5,1.0,0,0} 
\definecolor{ultramarine}{rgb}{0.07, 0.04, 0.56}
\definecolor{cadmiumgreen}{rgb}{0.0, 0.42, 0.24}
\definecolor{indigo(dye)}{rgb}{0.0, 0.25, 0.42}

\title[Signatures of metal-free star formation in Planck 2015 Polarization Data]{Signatures of metal-free star formation in Planck 2015 Polarization Data}

\author[Vinicius Miranda, Adam Lidz, Chen He Heinrich and Wayne Hu]{
Vinicius Miranda,$^{1}$\thanks{E-mail: vinim@sas.upenn.edu}
Adam Lidz,$^{1}$
Chen He Heinrich$^{2}$
and Wayne Hu$^{2,3}$
\\
$^{1}$Center for Particle Cosmology, Department of Physics and Astronomy, University of Pennsylvania, Philadelphia, PA 19104\\
$^{2}$Kavli Institute for Cosmological Physics,
Enrico Fermi Institute, University of Chicago, Chicago IL 60637 \\
$^{3}$Department of Astronomy \& Astrophysics,
 University of Chicago, Chicago IL 60637
}

\date{Accepted XXX. Received YYY; in original form ZZZ}
\pubyear{2016}

\begin{document}
\label{firstpage}
\pagerange{\pageref{firstpage}--\pageref{lastpage}}
\maketitle

\begin{abstract}
Standard analyses of the reionization history of the universe from Planck cosmic microwave background (CMB) polarization 
measurements consider only the overall optical depth to electron scattering ($\tau$), and further assume
a step-like reionization history. However, the polarization data contain information beyond the overall optical depth, and the assumption of a step-like function may miss high redshift contributions to the optical depth and lead to biased $\tau$ constraints. Accounting for its full reionization information content, we reconsider
the interpretation of  Planck 2015 Low Frequency Instrument (LFI) polarization data using simple, yet physically-motivated reionization models.
We show that these measurements still, in fact, allow a non-negligible
contribution from metal-free (Pop-III) stars forming in mini-halos of mass  $M \sim 10^5-10^6 M_\odot$ at $z \gtrsim 15$, provided this mode of star formation is fairly inefficient.
Our best fit model includes
an early, self-regulated phase of Pop-III star formation in which the reionization history has a gradual, plateau feature. In this model, $\sim$20\% of the volume of the universe is ionized by $z \sim 20$,
yet it nevertheless provides a good match to the Planck LFI measurements. Although preferred when the full information content of the data is incorporated, this model would spuriously be disfavored in the standard analysis. 
This preference is driven mostly by excess power from  E-mode
polarization  at multipoles of $10 \lesssim \ell  \lesssim 20$, which may reflect remaining systematic errors in the data, a statistical fluctuation, or signatures of the first stars. Measurements
from the Planck High Frequency Instrument (HFI) should be able to confirm or refute this hint
and future cosmic-variance limited E-mode polarization surveys can provide substantially
more information on these signatures. 
\end{abstract}


\section{Introduction} \label{sec:intro}

Measurements of the CMB polarization on large angular scales probe the probability that CMB photons scatter off of free electrons produced during the Epoch of Reionization (EoR) \citep{1982RSPTA.307...97H,1987MNRAS.227P..33E,1997PhRvD..55.1822Z}. The EoR is the time period during which the first stars, galaxies, and accreting black holes formed, emitted ultraviolet photons,
and ionized surrounding hydrogen and helium gas \citep{Loeb13}. 
Recent measurements from the Planck LFI and HFI suggest lower values for the optical depth to electron scattering than implied by earlier Wilkinson Microwave Anisotropy Probe (WMAP) observations \citep{2015arXiv150201589P,2016arXiv160503507P,2013ApJS..208...19H}. These new measurements have important implications for our understanding of the EoR, and have been studied extensively in the literature  \citep[e.g.][]{2015ApJ...802L..19R,2015MNRAS.454L..76M,Becker:2015lua,2016MNRAS.460..417S,2016arXiv160505374G}.

In general, previous theoretical studies have emphasized the lower values of the optical depth preferred by Planck. This is in broad agreement with direct censuses of high redshift galaxy populations -- which suggest fairly low ionizing emissivities -- and other reionization observables \citep{2015ApJ...802L..19R}.

However, it is still plausible that reionization is a very extended process, and that a small fraction of the  volume of the intergalactic medium (IGM) remains ionized out to rather high redshift. In this context, it is important to make use of the full information content of the CMB E-mode polarization power spectrum, which cannot be reduced to a single number quantifying the overall optical depth.  It is also important to ensure that CMB E-mode inferences on the overall optical depth are not biased by implicit assumptions on the reionization history itself.

Toward this end, we further consider the analysis of Planck LFI data in \cite{2016arXiv160904788H}. These authors use a principal component (PC) analysis methodology, in which 
any given reionization history between $6 < z < 30$  is characterized by five numbers.  These eigenmode
amplitudes completely capture the impact of reionization in this redshift range on the CMB polarization power spectrum observables \citep{Hu:2003gh,Mortonson:2007hq,Mortonson:2008rx}.
Interestingly, that study finds a two sigma preference for a contribution to the optical depth from $z \geq 15$ in the LFI data. 

This result is intriguing because it may indicate the existence of ionizing sources at very high redshift, and this runs somewhat counter to the current conventional wisdom that Planck requires late reionization. For example, this high redshift $\tau$ contribution may arise from metal-free (Pop-III) stars forming in minhalos ($M \sim 10^5-10^6 M_\odot$) at 
$z \sim 20$ \citep{2009Natur.459...49B}. In fact, previous work suggested that Planck data may be used to detect such signatures \citep{2012ApJ...756L..16A}, following-up in part on earlier work from e.g. \cite{2003ApJ...595....1H} that considered the high $\tau$ indicated by early WMAP data, but this possibility has 
not yet been investigated with the actual Planck data. The Pop-III mode of star formation at high redshift must be fairly inefficient, otherwise these sources would rapidly reionize the universe and overproduce the observed optical depth \citep{Visbal15}, but a small fraction of the gas in these early halos may nevertheless be converted into massive metal-free stars. 

To explore this scenario, we use the fast likelihood technique developed and tested in \cite{2016arXiv160904788H} to constrain simple reionization models. Although the PC methodology can not be used to reconstruct the reionization history directly, it is ideal for forward-modeling. This paper hence provides an illustrative example of how the PC technique can be used to constrain the parameters of reionization models. Software implementing the fast likelihood calculations of \cite{2016arXiv160904788H} is available upon request, and should provide a useful tool for reionization modelers.

\section{Reionization Models}
\label{sec:popII_popIII}

In the reionization models considered here, we assume that metal-free stars form exclusively in mini-halos that rely on molecular hydrogen for cooling, while we suppose that normal (Pop-II) star formation occurs in more massive halos where the gas cools by atomic line emission \citep[e.g.][]{Haiman06}. The rationale for this split is that atomic cooling halos likely have one or more mini-halo
progenitors that formed Pop-III stars. Supernovae from these short-lived stars subsequently enrich the surrounding gas with heavy elements and this shuts off the Pop-III star-formation mode before the atomic cooling ``descendent'' halos collapse.
The redshift evolution of the average fraction of ionized hydrogen, $\avg{x_i}$, is determined by the following ``photon-counting'' equation \citep{Shapiro87,Madau99,Loeb13}:
\begin{align}
\label{eq:dxdt}
\frac{d\avg{x_i}}{dt} = & \frac{d}{dt} \left( \zeta_{\rm II} f_{c,{\rm II}}   + \zeta_{\rm III} f_{c,{\rm III}}\right)  - \frac{\avg{x_i}}{\bar{t}_{\rm rec}(z)}. 
\end{align}
The first two terms describe the rate at which Pop-II and Pop-III stars ionize surrounding hydrogen atoms, while the final term accounts for recombinations. 
In our model, the ionizing photon production in each population of stars traces the rate at which matter collapses into their respective dark matter host halos. Hence, $f_{c,{\rm II}}$ denotes the collapse fraction in halos above the minimum mass required to host Pop-II stars $M_{\rm min,II}$, while $f_{c,{\rm III}}$
denotes
the collapse fraction in the lower mass halos  between $M_{\rm min,{\rm III}} \leq M \leq M_{\rm min,II}$ in which Pop-III stars reside. To compute these collapse fractions, we adopt the halo mass function of  \cite{Sheth:1999su}. 

The recombination term depends on the average time required for ionized hydrogen to recombine, $\bar{t}_{\rm rec}(z)$. This in turn depends on the temperature and the clumpiness of the ionized gas that resides in the intergalactic medium, wth $\bar{t}_{\rm rec} = C/[\alpha_B(T) \bar{n}_e(z)]$. Here $C$ is the clumping factor of the ionized gas,  $\alpha_B(T)$ is the recombination coefficient for gas at temperature $T$ and $\bar{n}_e(z)$ is the average electron number density at redshift $z$. We assume case-B recombination and evaluate
$\alpha_B$  at a redshift independent temperature of $T=2 \times 10^4$\,K \citep{Hui:1997dp}. We choose a redshift-independent clumping factor of $C=2$, in agreement with recent simulations that find small values for the clumping factor \citep{Pawlik:2008mr,McQuinn:2011aa}.

The minimum mass of the source host halos is set such that Pop-III stars form only in halos where molecular hydrogen cooling is efficient, while normal star formation is assumed
to take place in atomic cooling halos. On this basis, we follow \cite{Haiman06} and assume that metal-free stars form in halos with virial temperatures between $400 \,{\rm K} < T_{\rm vir} < 10^4$\,K, while
Pop-II stars form in halos with virial temperature above $10^4$\,K. The lower virial temperature is an optimistic value for efficient molecular hydrogen cooling and so we 
subsequently test the impact of raising the minimum host virial temperature to $10^3$\,K \citep{2003ApJ...592..645Y} (see \S \ref{likelihood}).
These considerations set the redshift-dependent values of the minimum host halo masses, $M_{\rm min,III}$, $M_{\rm min,II}$.

The ionizing efficiency parameters $\zeta_{\text{II}}$ and $\zeta_{\text{III}}$ can each be written as the product of several uncertain factors, with $\zeta = A_{\text{He}} \times N_\gamma \times f_{\ast} \times f_{\text{esc}}$ \citep{Loeb13}. Here, $N_\gamma$ is the number of ionizing photons produced per baryon converted into stars, $f_\ast$ is the star-formation efficiency -- i.e., it is the fraction of halo baryons that are converted into stars, while $f_{\text{esc}}$ is the fraction of ionizing photons that escape the host halo and ionize atoms in the IGM. Furthermore, $A_{\text{He}} = 4/(4 - 3Y_p)$ is a rescaling factor to account for (singly-ionized) helium and $Y_p$ is the primordial helium mass fraction. 
Here we assume that metal-free stars have large masses and high surface temperatures, and so produce copious numbers of ionizing photons: our fiducial Pop-III star model adopts $A_{\text{He}} \times N_{\gamma,\text{III}} = 40,000$ \citep{Bromm01,Schaerer02}. Note that the expected yield of ionizing photons depends on the Initial Mass Function for this mode of star formation \citep{2000ApJ...528L..65T}, which is still uncertain.
 In addition, numerical simulations of Pop-III star formation suggest that a high fraction of ionizing photons are able to escape from their host halos and ionize atoms in the IGM \citep{2006ApJ...639..621A}. Our model therefore takes $f_{\text{esc}, \text{III}}=0.5$. Finally, we vary the Pop-III star-formation efficiency over a broad range of values from $10^{-4} < f_{\ast,\text{III}} < 0.1$. 

\begin{figure}
\includegraphics[width=0.95\linewidth]{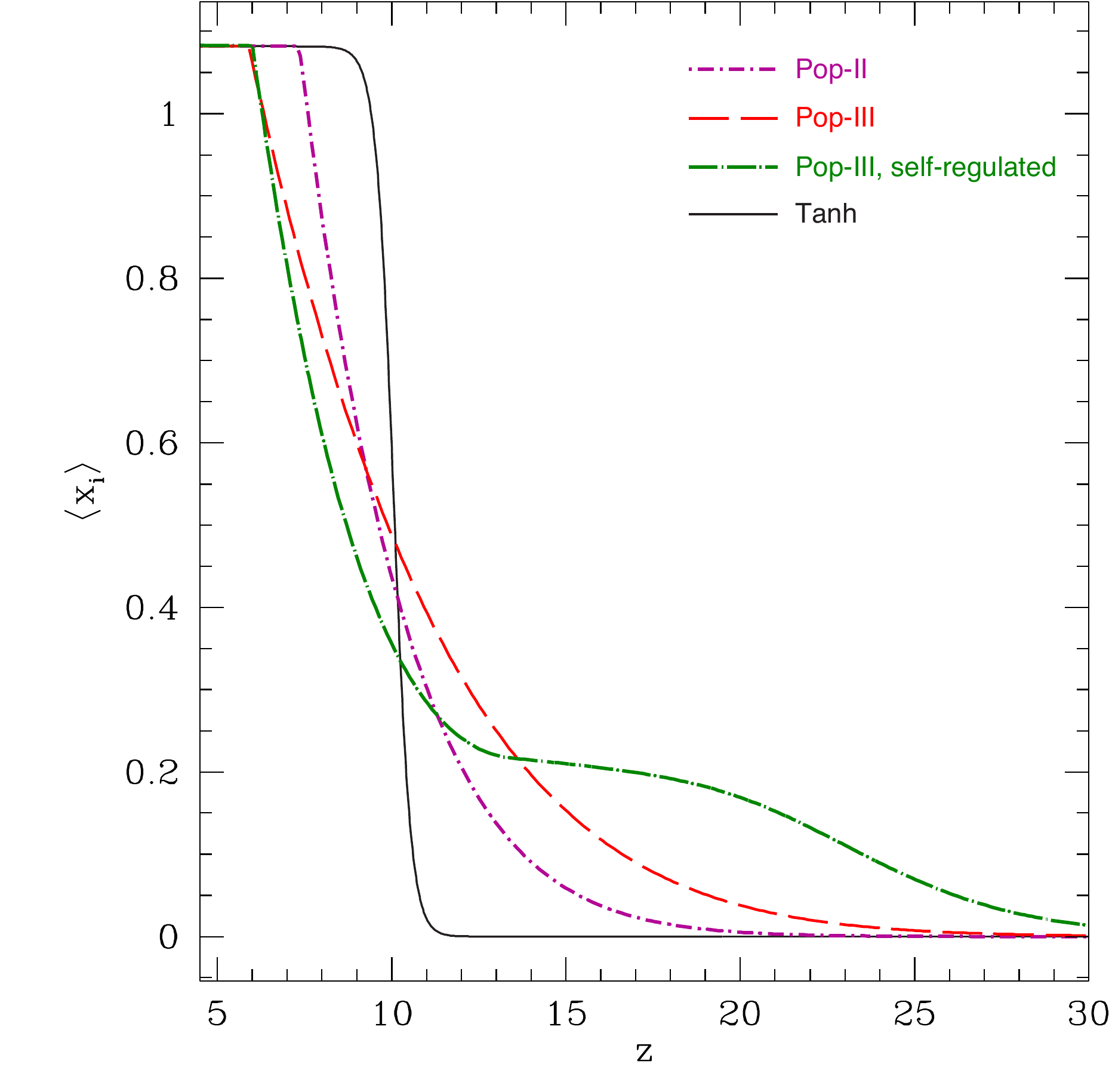}
\caption{Reionization history in the various models from fits to the Planck 2015 data with parameters from Tab.~\ref{tab:models}.   In order of increasing high redshift ionization contributions these are: step-like Tanh  model (black solid line),  model with only Pop-II stars  (purple dot-short-dashed line), model with additional Pop-III contributions (red long-dashed line) and where those contributions
are self-regulated (with $x_{\text{max}}=0.2$, green dot-dashed line). A broad range of ionization histories remain consistent with the Planck 2015 data. }
\label{fig:xe} 
\end{figure}

In the case of Pop-II stars in atomic cooling halos, the expected efficiency parameters take rather different values. In this case, typical values considered in the literature are $N_{\gamma,\text{II}} \sim 4,000$, $f_{\text{esc},\text{II}}=0.1$, $f_{\ast,\text{II}} = 0.1$, corresponding to $\zeta_{\text{II}} \sim 50$ \citep{Lidz16}.
Here, we conservatively allow the Pop-II star-formation efficiency coefficient to vary over the range of $5 < \zeta_{\text{II}} < 500$.  In Fig.~\ref{fig:xe} we show an example of a Pop-II dominated ionization model
and one where the Pop-III contribution dominates at high redshift. We also compare this history to the step-like
Tanh reionization history
that is assumed in the standard Planck analysis.   As we shall see in the next section, the parameters
of these models are chosen to best fit the Planck 2015 data and are listed in Tab.~\ref{tab:models}.
 The Tanh history has no ionization at high redshift
by assumption of form whereas the Pop-III model has a broader tail than the
Pop-II model though both have a sharp decline with redshift.

\begin{table}
\centering
\caption{Model Parameters and Fits}
\begin{tabular}{ l  c c  c  c }
\hline Model  & $\tau$            & $\zeta_{\text{II}}$ &    $ f_{\ast,\text{III}}$        &\!\!\!\!\!\! $ \chi_\text{model}^2-\chi_\text{Pop-II}^2  $ \\ \hline
Tanh & 0.079 & -- &  -- & 0.97 \\
Pop-II & 0.081 & 19.6& -- & 0\\
Pop-III & 0.089 & 6.91  & $0.00045 $ & -1.0\\
Pop-III self-reg. & 0.010 & 12.1 & $0.0011$ & -$2.2$\\
\end{tabular}
\label{tab:models}
\end{table}

The sharp decline of the Pop-III contribution is itself dependent on our model assumptions.
The Pop-III star formation mode is likely fragile, and disrupted by a range of feedback effects including that from dissociating ultraviolet radiation, photo-ionization feedback, supernova feedback, and chemical enrichment \citep[e.g.][and references therein]{2009Natur.459...49B}. In our fiducial Pop-III model, we assume that the impact of feedback is roughly encapsulated in 
 the efficiency parameter, $\zeta_{\text{III}}$. While an overall efficiency parameter may capture internal feedback, ``external feedback'' from surrounding sources is likely important as well; this should impact the overall progression of reionization, and is unlikely adequately captured by our single efficiency parameter.
 
A detailed modeling of feedback is well beyond the scope of this paper, but to explore its plausible impact we consider a second Pop-III model in which 
we multiply the Pop-III term in Equation~\ref{eq:dxdt} by $(1  -\avg{x_i}/x_{\text{max}})$ if $\avg{x_i}  \leq x_{\text{max}}$, while we set this to zero
when $\avg{x_i} > x_{\text{max}}$. This form is meant to mimic that 
Pop-III star formation may be ``self-regulating'' \citep{2012ApJ...756L..16A}.
 Here metal-free star formation is allowed to continue only in neutral regions, with an increasing suppression factor, while it is suppressed completely
when the average ionization fraction exceeds a threshold value of $x_{\text{max}}$. The suppression is meant to roughly capture two separate feedback effects. First, Pop-III star formation will
be truncated or less efficient in ionized regions where the gas is photo-heated and unable to collapse into mini-halos. Next, and more important, is that early Pop-III stars will produce
a dissociating ultraviolet radiation background that will prevent or suppress molecular hydrogen cooling and the formation of additional stars in mini-halos \citep{1997ApJ...476..458H}.

 Although this is only a toy model, we find that $x_{\text{max}}=0.2$ produces a plateau-feature in the reionization history similar to that seen in the simulations of \cite{2012ApJ...756L..16A},
 although our plateau is slightly more pronounced.
In this model, we also boost the ionizing efficiency factor to start reionization as early as possible to $A_{\text{He}} \times N_{\gamma,\text{III}} \times f_{\text{esc,III}} = 10^5$, although this is
of course degenerate with $f_{\ast,\text{III}}$. 
We show an example self-regulated Pop-III model in Fig.~\ref{fig:xe}.
If the reader prefers to compare models at fixed $A_{\text{He}} \times N_{\gamma,\text{III}} \times f_{\text{esc,III}} $ one can rescale the
self-regulated star formation efficiency, $f_{\ast,\text{III}}$ upwards by a factor of five. 
We show below that this model is a marginally better fit to the Planck LFI data. 

We can further consider the plausibility of our model value, $x_{\text{max}} \sim 0.2$, by calculating the average specific intensity of the dissociating ultraviolet background at the redshifts
and ionized fractions of interest. Ultraviolet photons in the portion of the Lyman-Werner band between
$11.2\,{\rm eV} \le h\nu \le 13.6\,{\rm eV}$ can dissociate molecular hydrogen, while even the pre-reionization IGM is largely transparent to such photons, since they lie beneath the hydrogen photoionization edge.
However, below the ionization edge photons can still be absorbed out of the dissociating background in Lyman series lines from neutral hydrogen atoms: this imprints a sawtooth feature on the spectrum of the dissociating background \citep{1997ApJ...476..458H}.

A rough estimate of the average specific intensity of dissociating ultraviolet radiation in the Lyman-Werner band, $J_{\text{LW}}$, in our model can be made according to  \citep[e.g.][]{Visbal15}:
\begin{align}
J_{\text{LW}}(z) =  \frac{c}{4\pi} \frac{\bar{\rho}_b(z)}{m_p} f_{\ast,\text{III}} \frac{N_\text{LW} h \nu_\text{LW}}{\Delta \nu_{\text{LW}}} 
& \left[f_{c,\text{III}}(z) - f_{c,\text{III}}(z_{\text{hor}})\right].
\label{eq:jlw}
 \end{align}
Here $c$ is the speed of light, $\bar{\rho}_b(z)/m_p$ is the baryonic number density at redshift $z$, and $f_{\ast,\text{III}}$ is the Pop-III star formation efficiency, while $N_{\text{LW}}$ is the number of 
dissociating Lyman-Werner photons per baryon converted into stars, $h \nu_{\text{LW}}$ is their typical
energy, and $\Delta \nu_{\text{LW}}$ is the bandwidth of such photons. As in \cite{Visbal15}, we assume that the Lyman-Werner background at redshift $z$ is set by sources out to
$z_{\text{hor}} = 1.015 z$. This ``Lyman-Werner (LW) horizon'' redshift, $z_\text{hor}$, reflects the typical distance a Lyman-Werner photon travels before redshifting into a Lyman series line and being
absorbed out of the background. Equation~\ref{eq:jlw} therefore ignores any absorption between $z$ and $z_\text{hor}$ and neglects any contribution from sources beyond this horizon; it is intended
to capture an average suppression from the Lyman series absorption without tracking the full sawtooth effect.
In this estimate, we have ignored the model truncation of Pop-III sources in ionized regions and from the build-up of dissociating radiation. We have also neglected any Pop-II contribution, which should be a good approximation for present purposes.
We adopt $N_{\text{LW}}=3400$ and $\nu/\Delta \nu_{\text{LW}} = 4.9$ following \cite{2014MNRAS.445..107V}.  
According to the estimate of Equation~\ref{eq:jlw}, the intensity of dissociating radiation reaches values of $J_{\text{LW}}/J_{21} \sim 0.1-0.2$ -- where $J_{21}$ is the specific intensity in units of $10^{-21}$ ergs cm$^{-2}$ s$^{-1}$ Hz$^{-1}$ str$^{-1}$ --
at $z \sim 25$ for $f_{\ast,\text{III}} =10^{-3}$. For this redshift and efficiency factor, $\avg{x_i} \sim 0.1$. Since the contribution of Pop-III stars is suppressed in our self-regulated model
by a factor of $1 - \avg{x_i}/x_\text{max}$, this amounts to a factor of two at this redshift and stage of reionization. In other words, our model reduces the contribution of Pop-III stars
by a factor of two at $J_\text{LW}/J_{21}\sim 0.1-0.2 $.
Note that although $J_\text{LW}$ at a given redshift depends on $f_{\ast,\text{III}}$, the intensity at a given $\avg{x_i}$ is insensitive to the star-formation efficiency since both $J_\text{LW}$ and $\avg{x_i}$ scale in proportion to $f_{\ast,\text{III}}$.

This is a bit on the high side of the threshold specific intensities at which previous studies suggest that Lyman-Werner background photons will largely suppress molecular hydrogen cooling: for example,
\cite{2012ApJ...756L..16A} quote plausible threshold values of $J_{\text{LW,th}}/J_{21}= 0.01-0.1$  (see also the references in that work). This may partly reflect that Equation~\ref{eq:jlw} only provides a rough estimate: \cite{2012ApJ...756L..16A} find a broadly similar plateau
feature at a comparable ionized fraction, although they adopt a lower dissociation threshold. 

In any case, it is likely that any ionization plateau is less well-defined than in our toy model. For one, larger mass halos will generally contain more molecular gas and more intense ultraviolet radiation should therefore be required to suppress cooling in such halos. For example, in the analytic model of \citet{Visbal15}, the authors self-consistently model the build-up of the average Lyman-Werner background along with the ionization history, while incorporating a halo-mass dependent ultraviolet background threshold (above which cooling is completely suppressed). Their model ionization
histories do not include a plateau feature, in contrast to the study of \cite{2012ApJ...756L..16A}, which adopts a mass-independent threshold. In addition,
some halos will turn around early and be largely self-shielded before the ultraviolet background is intense enough to dissociate molecular hydrogen, while the ultraviolet background will itself
be somewhat inhomogeneous.\footnote{Note, however, that the study of \cite{2012ApJ...756L..16A} does model inhomogeneities in the Lyman-Werner background.} The mass dependence and inhomogeneities may act to soften any plateau feature. 
Significantly more detailed models than considered here will be required to understand these issues better.

\begin{figure}
\includegraphics[width=0.95\linewidth]{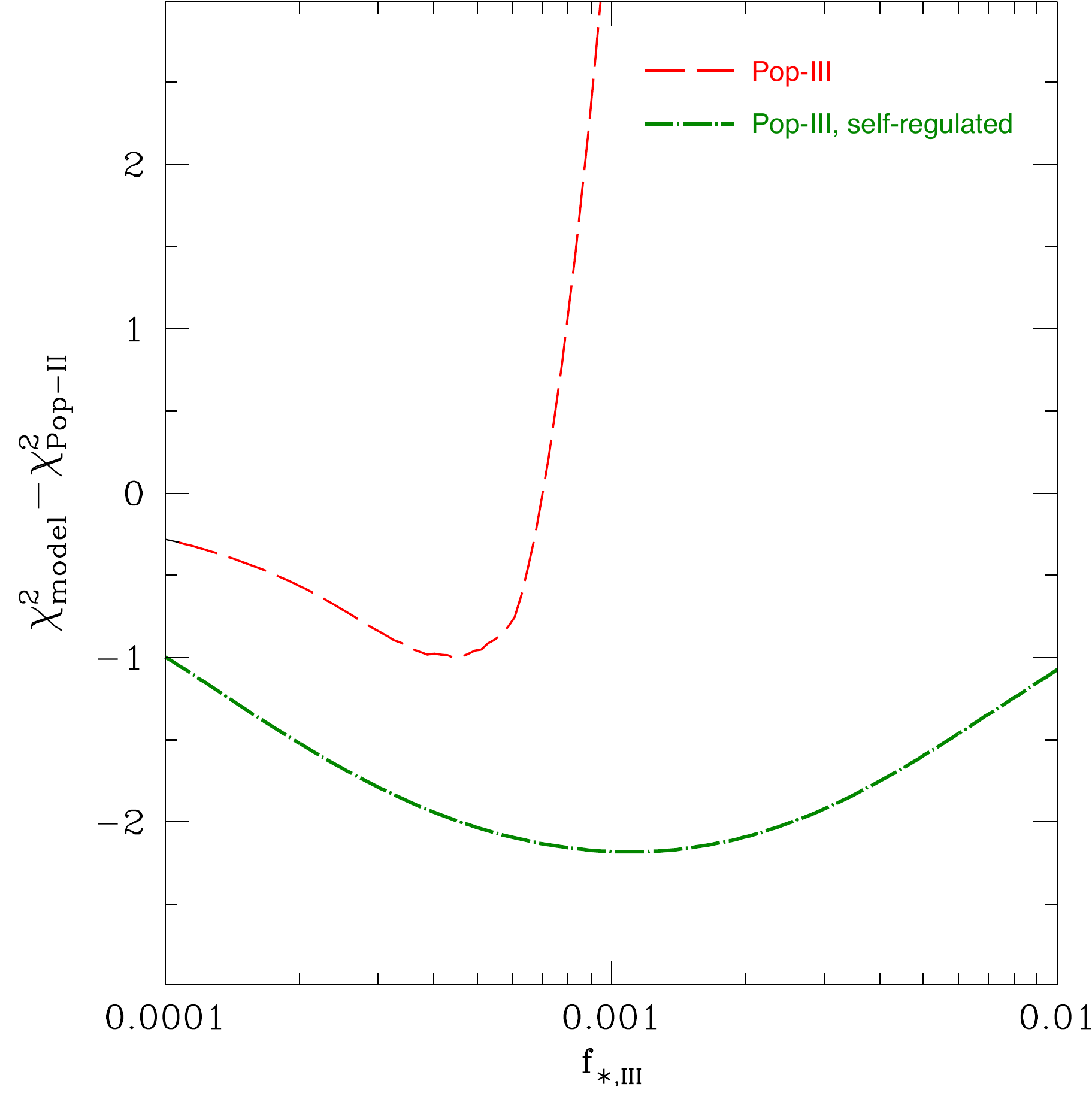}
\caption{The effective $\chi^2$ of various  Pop-III models to the Planck LFI data relative to $\chi_\text{Pop-II}^2$ for  the
best fit model with Pop-II stars only. The red dashed line shows $\chi^2$ as a function of the Pop-III star formation efficiency parameter $f_{\ast,\text{III}}$ for the fiducial model after minimization
 with respect to the Pop-II efficiency parameter.  The green dot-dashed line shows the same for our self-regulated Pop-III model.}  
\label{fig:posterior} 
\end{figure}

\section{Planck 2015 Analysis}

\label{likelihood}

In order to compare these models with Planck 2015 data, we use the complete analysis of the
reionization information from \cite{2016arXiv160904788H}. Each reionization history is projected onto a basis of
principal components $S_a(z)$ of a cosmic variance limited polarization measurement around some fixed
but otherwise arbitrary fiducial ionization history  $\langle x^{\text{fix}}_i\rangle(z)$.  Only modes $a=1,\ldots, 5$  are
needed to specify the entire information content of the E-mode polarization
power spectrum for $6 < z < 30$ \citep{Hu:2003gh}. For any given ionization history, $\avg{x_i}(z)$, the amplitude of each mode is  determined according to:
\begin{equation}
m_{a}=
\frac{1}{24}  \int _{6}^{30} dz\, {S_{a}(z) [\langle x_i \rangle(z)-\langle x^{\text{fix}}_i\rangle(z)]}.
\label{eq:xetommu}
\end{equation}
  We assume that $\langle x_i \rangle(z)=\langle x^{\text{fix}}_i\rangle(z)$ outside
of the range of the integral.   The fixed ionization history is specified in  \cite{2016arXiv160904788H} such that hydrogen is fully ionized and that helium is singly ionized at $z \le 6$ and helium is fully ionized
 at $z \lesssim 3.5$.  At $z\ge 30$, the ionization history returns to that given by recombination. 
 
Given $m_a$, the 5 principal component amplitudes of a model,
we evaluate the effective likelihood
${\cal L}(m_a)$
 of that model using the technique of~\cite{2016arXiv160904788H}. 
Here we interpret this as an effective $\chi^2 = -2\ln {\cal L}$.   In order
to set the baseline values for comparison, we first minimize $\chi^2$
without any Pop-III contribution to find $\chi^2_{\text{Pop-II}}$.  This model is specified as ``Pop-II'' in 
Tab.~\ref{tab:models} and is itself a better fit than the best fit Tanh model.
Likewise, for each Pop-III model parameterized by the
 Pop-III star formation efficiency parameter $f_{\ast,\text{III}}$, we minimize $\chi^2$ over 
the Pop-II efficiency parameter, $\zeta_\text{II}$.  
Note that this minimization is equivalent to marginalization over $\zeta_\text{II}$ if
the joint posterior probability distribution is a multivariate Gaussian.

\begin{figure}
\includegraphics[width=0.95\linewidth]{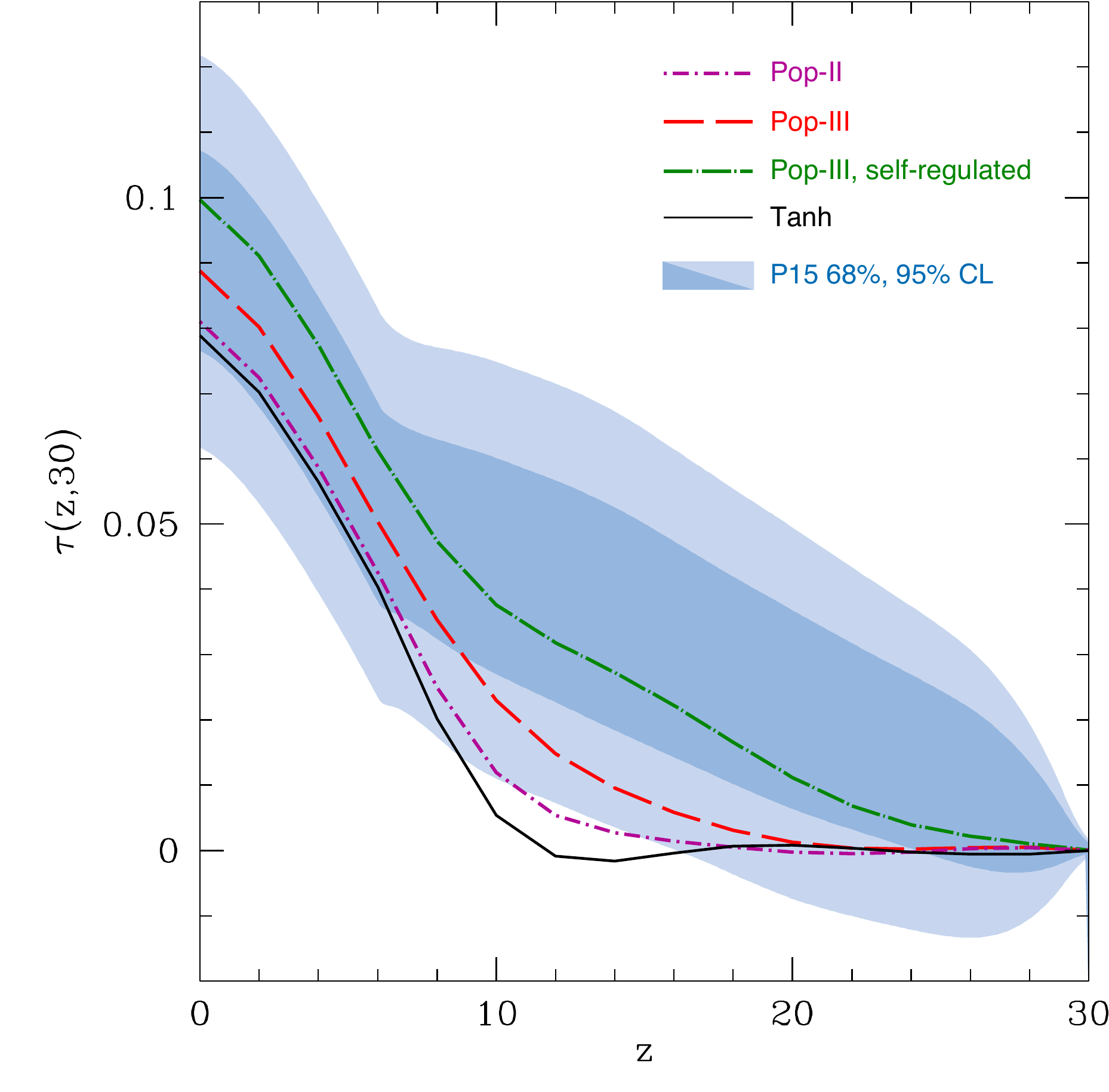}
\caption{Cumulative optical depth $\tau(z,30)$ in the Planck 2015 analysis. Blue shaded regions are the $68\%$ and $95\%$ constraints from the complete PC analysis. This is compared with the best fit models from Tab.~\ref{tab:models}:  
Tanh    (black solid line),
Pop-II only  (purple dot-short-dashed line), additional Pop-III fiducial (red long-dashed line),
additional Pop-III self-regulated   (green dot-dashed line) models.}
\label{fig:cum_tau} 
\end{figure}

 In Fig.~\ref{fig:posterior}, we show
the difference $\chi^2_{\rm model} - \chi^2_{\text{Pop-II}}$ as a function of $f_{\ast,\text{III}}$
 (with $A_{\text{He}} \times N_\gamma \times f_{\text{esc}}=2 \times 10^4$ in the ``Pop-III'' fiducial model and $A_{\text{He}} \times N_\gamma \times f_{\text{esc}}=10^5$ in the ``Pop-III, self-regulated'' model).   The model at the minimum of this curve
 for each case is given in Tab.~\ref{tab:models}.
Let us consider first the fiducial model (red dashed line, Figure~\ref{fig:posterior}). This shows that including Pop-III stars in our fiducial model with low efficiency, $f_{\ast,{\rm III}} \sim$ a few $\times$ $10^{-4}$ only
slightly improves the fit to the Planck data, with $\Delta \chi^2\approx1$. Furthermore, in qualitative agreement with previous work \citep{Visbal15,2016MNRAS.460..417S}, if Pop-III stars form with too great an efficiency they overproduce the E-mode polarization power
and so the Planck data imply an interesting upper limit on the efficiency of metal-free star formation in minhalos. 

However, our fiducial model assumes a redshift-independent efficiency factor which is too simplistic, as we discussed previously. In our self-regulated Pop-III model, including metal-free star formation improves the fit further with $\Delta \chi^2 =2.2$ compared to the best-fit model with no Pop-III component (see the green dot-dashed line in the figure). In this case, the preferred efficiency factor is $f_{\ast,{\rm III}} \sim 10^{-3}$ which corresponds to roughly one $\sim 100 M_\odot$ star per halo of mass
$M \sim 6 \times 10^5 M_\odot$ (fairly typical of the mini-halos), before this mode of star-formation is shutoff (completely at $\avg{x_i} =0.2$ in this model).\footnote{Note again that we assume a higher
ionizing efficiency in the self-regulated model. If one prefers the lower efficiency assumed in our fiducial model, one can rescale the star-formation efficiency in the self-regulated model upwards
by a factor of five.}

Although the preference for a  Pop-III contribution is weak -- and we are not claiming that the Planck data demand these sources -- 
the more important point here is that models with significant high redshift contributions to the ionization history are still viable.  The ionization history for these best fit models are shown in Figure~\ref{fig:xe}. 
In the self-regulated Pop-III model  the extended ionization plateau at $\avg{x_i} \sim 0.2$ near $z \sim 20$ is strikingly different than the best fit Pop-II model, in which the ionization fraction is negligible at such high redshifts. Furthermore even the Pop-II only model has a more extended high redshift tail than the Tanh model upon which the standard Planck analysis of the overall optical 
depth $\tau$ is based.  Tab.~\ref{tab:models} also gives $\tau$ for the various models which are higher
for those with high redshift contributions.

\begin{figure}
\includegraphics[width=0.95\linewidth]{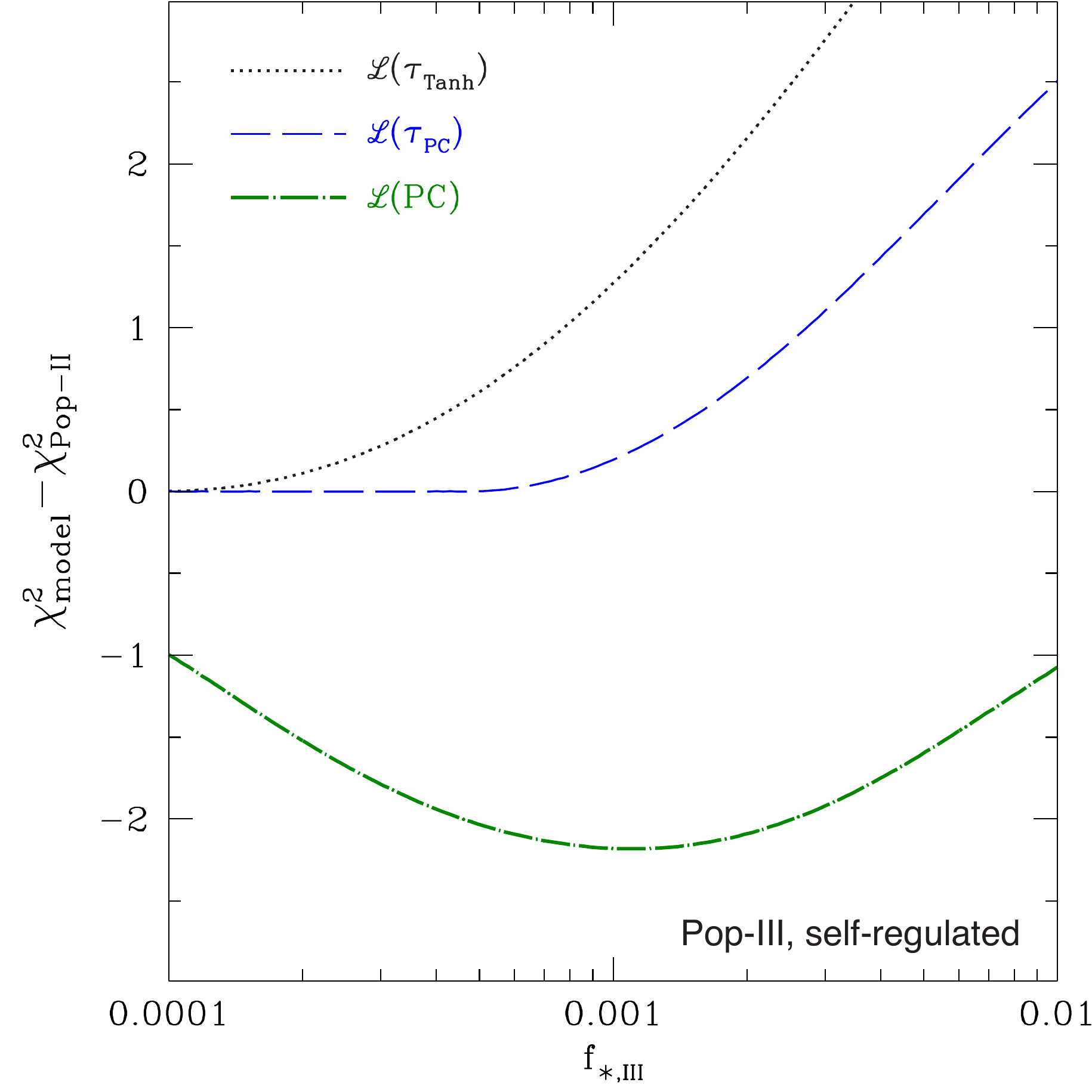}
\caption{Impact of incorrectly or incompletely assessing the reionization information content of Planck 2015 LFI data.   Effective $\chi^2$ as in Fig.~\ref{fig:posterior} for the self-regulated Pop-III model
but employing the likelihood from: i) the standard Tanh $\tau$ constraint (black dotted line), which is 
incorrect in the Pop-III context; ii) the overall $\tau$ constraint from the PC analysis (blue dashed line), which is
incomplete;  iii) the full  PC analysis from Fig.~\ref{fig:posterior} (green dot-dashed line) which reveals the  preference
for the model.  }  
\label{fig:posterior2} 
\end{figure}

In order to interpret these changes in the overall optical depth, it is useful to compute the cumulative optical depth
between redshift $z$ and $30$, $\tau(z,30)$. 
Rather than directly compute this from $\langle x_i \rangle(z)$ and cosmological parameters,
we  sum over contributions from each component $\tau^a(z,30)$
\begin{equation}
\tau(z,30) = \tau^{\rm fix}(z,30) + \sum_{a=1}^{5} m_a \tau^a(z,30).
\end{equation}
This provides a smoothed representation of the cumulative optical depth that can be
compared directly to the constraints derived in 
 \cite{2016arXiv160904788H}.    To evaluate $\tau^a(z,30)$ and the contribution from
 the fixed ionization history $\tau^{\rm fix}(z,30)$, we take the cosmological
 parameters
${\Omega_b h^2}=0.02224$, ${\Omega_m h^2}=0.1426$, and ${Y}_p=0.24534$.  

In Figure~\ref{fig:cum_tau}, we compare the best fit models to the constraints on the cumulative optical depth from  \cite{2016arXiv160904788H}. This gives some intuition as to what is driving the slight preference in the Planck LFI data for high redshift contributions from, for example, Pop-III stars. 
As discussed in \cite{2016arXiv160904788H}, the data favor a non-negligible contribution to $\tau(15, 30)=0.033\pm 0.016$. The reionization histories are not extended enough in the  Tanh  or the Pop-II only models to match the central value. In the case of the Pop-II only model, one needs to wait until lower redshift for the atomic cooling halos to collapse. If one supposes that the efficiency is extremely high in the atomic cooling halos, one can get a slightly earlier start in these models, but in this case reionization generally completes too early and one overproduces $\tau$. 

On the other hand, the Pop-III contribution helps extend reionization out to higher redshift since the molecular hydrogen cooling mini-halos collapse at higher redshift. In our redshift-independent Pop-III efficiency model, this contribution only helps slightly, however. In the self-regulated model, one gets a larger early contribution from Pop-III stars -- since the formation of these sources is truncated
early, the ones that form early can do so at higher -- yet still reasonable -- efficiency without overproducing $\tau$. In the self-regulated model with $x_{\text{max}}=0.2$, the 
model $\tau(z,30)$ always lies within the  $68\%$ confidence band preferred by the Planck data. In principle, the Pop-II component could be more extended than in
our model, which assumes a redshift independent efficiency factor, but the atomic cooling halos are rare at high redshifts and the efficiency factor would need to be uncomfortably large 
for these sources to have a non-negligible impact at $z \sim 20$, for example.

Figure~\ref{fig:cum_tau} also illustrates a final important point.   Note that the constraint on 
the overall optical depth $\tau(0,30)=0.092\pm 0.015$ is shifted upwards from that inferred
assuming the Tanh reionization history $\tau=0.079\pm 0.017$ \citep{2016arXiv160904788H}.   The latter does not correspond to
the overall optical depth in these Pop-III models and so applying this constraint would
spuriously disfavor such models.
 This point is made explicit in Figure~\ref{fig:posterior2}, which shows the likelihood for the Pop-III component using the Tanh $\tau$ constraint, rather
than our PC methodology. Clearly this would disfavor our best-fit model (at $\sim$1-$\sigma$) and rule-out many other viable Pop-III models. Using the overall $\tau(z=0,z_{\rm max})$ from the PC analysis restores
compatibility but the small preference for the self-regulated model is only revealed in the full PC analysis (green dot-dashed line). Correspondingly only when one makes use of the full information content
of the Planck 2015 data is the hint for high redshift ionization  apparent. Incorporating the full PC analysis, the best fit self-regulated model has a notably better $\chi^2$ than expected for the same
model using the Tanh $\tau$ constraint: the fit improves by $\Delta \chi^2 \sim 3.2$. 
Although this improvement does not represent a detection given the number of additional parameters,
the best fit Pop-III models are clearly still allowed.

\begin{figure}
\includegraphics[width=0.95\linewidth]{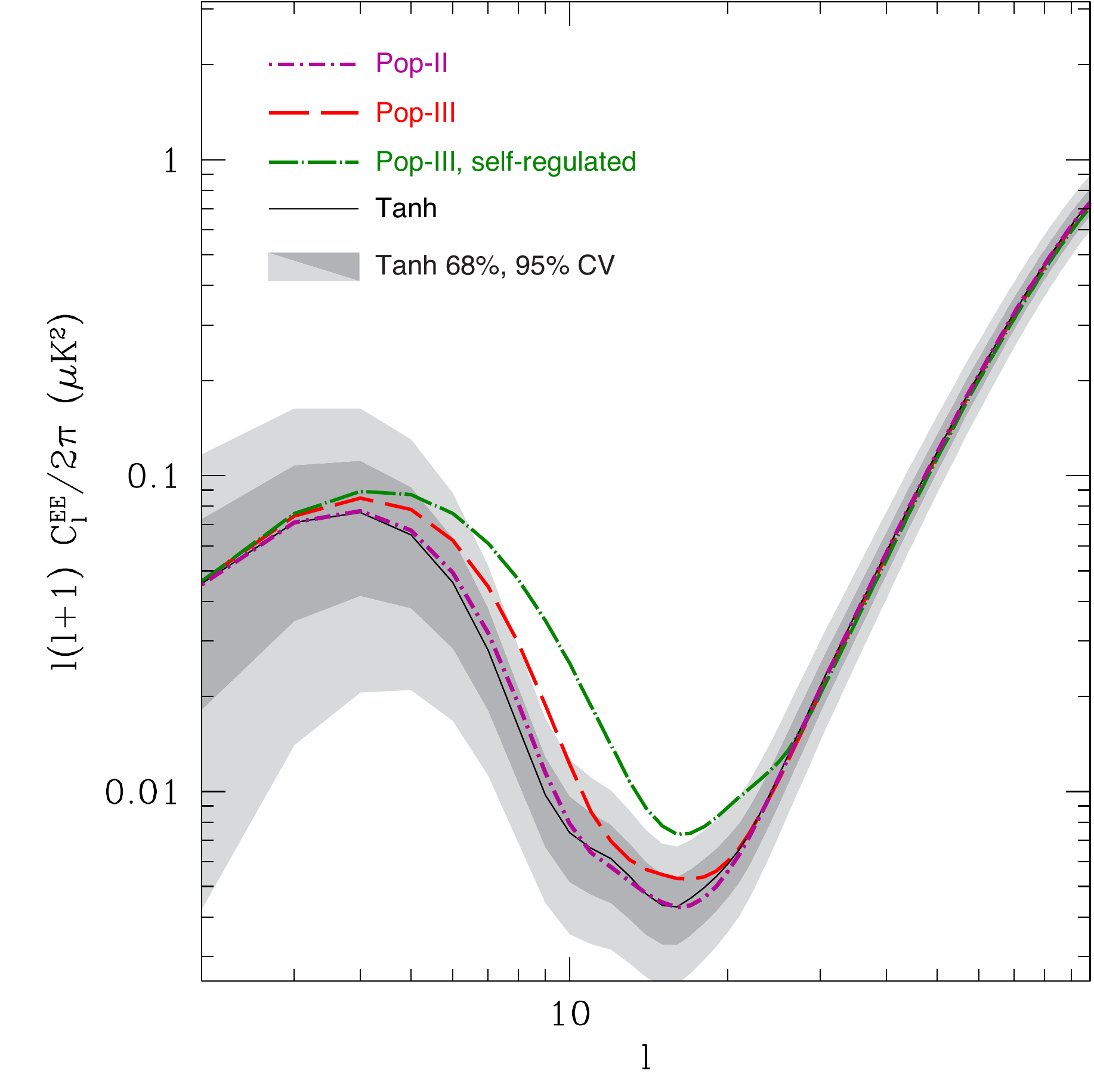}
\caption{E-mode polarization power spectra, $C_{\ell}^{\text{EE}}$, of the various best fit models as
in Fig.~\ref{fig:cum_tau}.
 The black solid line and gray shaded regions show the Tanh model, along with the $68\%$ and $95\%$ confidence regions expected in the cosmic variance (CV) limit.   In particular the two Pop-III models should be clearly
 distinguishable from the step-like Tanh model for a CV limited measurement.}     
\label{fig:clee}   
\end{figure}

The origin of these different inferences can be traced back to the E-mode power spectrum.
In Figure~\ref{fig:clee}, we show the power spectra of the best fit models. The most striking difference between the model power spectra is the excess power in the self-regulated Pop-III model at $10 \lesssim \ell  \lesssim 20$ when compared to the other models. This is the main feature of the Planck 2015 LFI data
 that drives the preference for this model. If one tries to generate excess power in a  step-like Tanh model at $10 \lesssim \ell \lesssim 20$, one will overproduce the power at lower multipoles and so high redshift contributions to the
 ionization history are effectively hidden from the usual constraints \citep{2016arXiv160904788H}. 
 
 This excess may reflect remaining systematic errors in the Planck 2015 LFI data from, for example, residual foregrounds. Alternatively,
 it may be an interesting signature of Pop-III star formation or other high redshift sources of ionization.
 Ultimately, if systematics are under control, the low $\ell$ polarization can be measured to the cosmic variance (CV) limit.  
 The shaded regions in Figure~\ref{fig:clee} show that if the true model is the Tanh one, CV limited measurements would test the Pop-III models at high significance.   This may be of interest for future space-based polarization missions such as the Cosmic Origins Explore (COrE) satellite \citep{2011arXiv1102.2181T}.

 Finally, we briefly comment on the sensitivity of our results to some of our model assumptions. First, we also explored self-regulated Pop-III models where we raised the minimum host
 virial temperature to $T_\text{vir}=10^3$\,K. After maximizing over the efficiency parameters in this model, the resulting ionization history looks nearly identical to the best fit in
 Figure~\ref{fig:xe}, which adopts $T_\text{vir}=400$\,K. However, the best-fit star formation efficiency goes up to $f_{\ast,\text{III}} = 0.0023 $: in this model, the higher star formation efficiency compensates for the boosted minimum host mass. Note also that we have ignored the effect pointed out in \cite{2010PhRvD..82h3520T}: while this should lead to interesting spatial variations, recent
 studies suggest a fairly small average suppression for star formation in mini-halos \citep{2012MNRAS.424.1335F}, and this may be compensated again by boosting our star formation efficiency parameter.  Lastly, we investigated a range of values of $x_\text{max}$. While slightly lower values of $x_\text{max}$ still match the data, the star formation efficiency needs
 to be increased in this case since the Pop-III phase is briefer in these models.

\section{Discussion}
\label{sec:discussion}

We have given an example illustrating how one may reach qualitatively different conclusions regarding the reionization history of the universe and the nature of the ionizing sources, when one accounts for the full information content of the Planck 2015 LFI data, rather than assuming a step-like ionization history.
For example, our best fit ionization history has $\avg{x_i}=0.2$ at $z \sim 20$: contrary to conventional wisdom, that data still allow an extended tail of ionization out to high redshift and a non-negligible contribution from metal-free stars in minihalos.
It will be interesting to see if the hint for high redshift contributions to $\tau$ is sharpened by future CMB polarization data. We eagerly await upcoming results from the new Planck
HFI large scale polarization data -- which is still proprietary -- to see if these observations strengthen the hint seen in the LFI data. Alternatively, these may reflect remaining systematics in the LFI data that do not also apply to the HFI data.
Based on Figure~\ref{fig:clee}, we expect that a future cosmic-variance limited experiment should be able to confirm or refute the high redshift contribution to $\tau$, although we postpone
a more detailed investigation of these prospects to future work. 

It will be extremely challenging to test the hint for non-negligible ionization fractions at $z \gtrsim 15$ by other means. The best alternative is likely redshifted 21 cm surveys but these
face challenges at high redshift owing to the bright galactic emission at the frequencies of interest. However, these surveys may find that the average neutral fraction rises more slowly towards high
redshift than expected in models with Pop-II stars alone (see Figure~\ref{fig:xe}). There may be differences in the sizes of the ionized regions in these scenarios as well, which may provide a potential signature \citep{2016arXiv160904400K}. However, the possibility of a significant contribution from Pop-III stars at high redshift may also somewhat
complicate the interpretation of the 21 cm power spectrum measurements \citep[e.g.][]{2013MNRAS.432.2909F}. Another possibility may be to extract signatures of early phases of patchy reionization using the Kinetic Sunyaev-Zel'dovich effect, but these will have to push to smaller angular scales which is challenging given foreground contamination. An extended self-regulated phase likely evades current constraints \citep{2012ApJ...756...65Z,2015ApJ...799..177G} on the
duration of reionization \citep{2013ApJ...769...93P}.

Finally, it is worth commenting on the implications of our findings for the goal of determining $\tau$ from the ionization history inferred from redshifted 21 cm observations  \citep{2016PhRvD..93d3013L}. This is an appealing idea, because the optical depth is an important nuisance parameter that can limit, for instance, inferences regarding the sum of the neutrino masses from upcoming CMB lensing measurements. However, this prospect becomes difficult if there are significant high redshift contributions to the optical depth, as hinted at in the Planck LFI data, which will be hard to extract from upcoming redshifted 21 cm surveys.

We hope that this work will encourage reionization modelers to adopt this PC analysis methodology
of \cite{Hu:2003gh} and effective likelihood of \cite{2016arXiv160904788H}. Here we have considered rather simple models for the ionization history in an effort to explore the broad-brush implications of the Planck data, but it would be interesting to explore more detailed reionization models and to combine the full information content of the CMB data with other reionization observables. 
It will also be interesting to explore whether dark matter annihilations \citep{2015arXiv151200526K} or early accreting black holes \citep{2004MNRAS.352..547R} provide interesting alternative sources of high redshift ionization to the Pop-III stars considered here.

 \medskip
 \noindent {\it Acknowledgments}:   We thank  Tom Abel, James Aguirre, and Kyungjin Ahn  for
 useful discussions.  WH thanks the Aspen Center for Physics, which is supported by National Science Foundation grant PHY-1066293, 
where part of this work was completed. CH and WH were supported by NASA ATP NNX15AK22G, U.S.~Dept.\ of Energy contract DE-FG02-13ER41958, and  the Kavli Institute for Cosmological Physics
at the University of Chicago through grants NSF PHY-0114422
and NSF PHY-0551142. Computing resources were provided by the University of Chicago Research Computing Center.  VM was supported in part by the Charles E.~Kaufman Foundation, a supporting organization of the Pittsburgh Foundation.

\vfill

\bibliographystyle{mnras}
\bibliography{ReiPopIII}

\begin{thebibliography}{}
\makeatletter
\relax
\def\mn@urlcharsother{\let\do\@makeother \do\$\do\&\do\#\do\^\do\_\do\%\do\~}
\def\mn@doi{\begingroup\mn@urlcharsother \@ifnextchar [ {\mn@doi@}
  {\mn@doi@[]}}
\def\mn@doi@[#1]#2{\def\@tempa{#1}\ifx\@tempa\@empty \href
  {http://dx.doi.org/#2} {doi:#2}\else \href {http://dx.doi.org/#2} {#1}\fi
  \endgroup}
\def\mn@eprint#1#2{\mn@eprint@#1:#2::\@nil}
\def\mn@eprint@arXiv#1{\href {http://arxiv.org/abs/#1} {{\tt arXiv:#1}}}
\def\mn@eprint@dblp#1{\href {http://dblp.uni-trier.de/rec/bibtex/#1.xml}
  {dblp:#1}}
\def\mn@eprint@#1:#2:#3:#4\@nil{\def\@tempa {#1}\def\@tempb {#2}\def\@tempc
  {#3}\ifx \@tempc \@empty \let \@tempc \@tempb \let \@tempb \@tempa \fi \ifx
  \@tempb \@empty \def\@tempb {arXiv}\fi \@ifundefined
  {mn@eprint@\@tempb}{\@tempb:\@tempc}{\expandafter \expandafter \csname
  mn@eprint@\@tempb\endcsname \expandafter{\@tempc}}}

\bibitem[\protect\citeauthoryear{{Ahn}, {Iliev}, {Shapiro}, {Mellema}, {Koda}
  \& {Mao}}{{Ahn} et~al.}{2012}]{2012ApJ...756L..16A}
{Ahn} K.,  {Iliev} I.~T.,  {Shapiro} P.~R.,  {Mellema} G.,  {Koda} J.,   {Mao}
  Y.,  2012, \mn@doi [\apjl] {10.1088/2041-8205/756/1/L16}, \href
  {http://adsabs.harvard.edu/abs/2012ApJ...756L..16A} {756, L16}

\bibitem[\protect\citeauthoryear{{Alvarez}, {Bromm}  \& {Shapiro}}{{Alvarez}
  et~al.}{2006}]{2006ApJ...639..621A}
{Alvarez} M.~A.,  {Bromm} V.,   {Shapiro} P.~R.,  2006, \mn@doi [\apj]
  {10.1086/499578}, \href {http://adsabs.harvard.edu/abs/2006ApJ...639..621A}
  {639, 621}

\bibitem[\protect\citeauthoryear{Becker, Bolton  \& Lidz}{Becker
  et~al.}{2015}]{Becker:2015lua}
Becker G.~D.,  Bolton J.~S.,   Lidz A.,  2015, \mn@doi [Publ. Astron. Soc.
  Austral.] {10.1017/pasa.2015.45}, 32, 45

\bibitem[\protect\citeauthoryear{{Bromm}, {Kudritzki}  \& {Loeb}}{{Bromm}
  et~al.}{2001}]{Bromm01}
{Bromm} V.,  {Kudritzki} R.~P.,   {Loeb} A.,  2001, \mn@doi [\apj]
  {10.1086/320549}, \href {http://adsabs.harvard.edu/abs/2001ApJ...552..464B}
  {552, 464}

\bibitem[\protect\citeauthoryear{{Bromm}, {Yoshida}, {Hernquist}  \&
  {McKee}}{{Bromm} et~al.}{2009}]{2009Natur.459...49B}
{Bromm} V.,  {Yoshida} N.,  {Hernquist} L.,   {McKee} C.~F.,  2009, \mn@doi
  [\nat] {10.1038/nature07990}, \href
  {http://adsabs.harvard.edu/abs/2009Natur.459...49B} {459, 49}

\bibitem[\protect\citeauthoryear{{Efstathiou} \& {Bond}}{{Efstathiou} \&
  {Bond}}{1987}]{1987MNRAS.227P..33E}
{Efstathiou} G.,  {Bond} J.~R.,  1987, \mn@doi [\mnras]
  {10.1093/mnras/227.1.33P}, \href
  {http://adsabs.harvard.edu/abs/1987MNRAS.227P..33E} {227, 33P}

\bibitem[\protect\citeauthoryear{{Fialkov}, {Barkana}, {Tseliakhovich}  \&
  {Hirata}}{{Fialkov} et~al.}{2012}]{2012MNRAS.424.1335F}
{Fialkov} A.,  {Barkana} R.,  {Tseliakhovich} D.,   {Hirata} C.~M.,  2012,
  \mn@doi [\mnras] {10.1111/j.1365-2966.2012.21318.x}, \href
  {http://adsabs.harvard.edu/abs/2012MNRAS.424.1335F} {424, 1335}

\bibitem[\protect\citeauthoryear{{Fialkov}, {Barkana}, {Visbal},
  {Tseliakhovich}  \& {Hirata}}{{Fialkov} et~al.}{2013}]{2013MNRAS.432.2909F}
{Fialkov} A.,  {Barkana} R.,  {Visbal} E.,  {Tseliakhovich} D.,   {Hirata}
  C.~M.,  2013, \mn@doi [\mnras] {10.1093/mnras/stt650}, \href
  {http://adsabs.harvard.edu/abs/2013MNRAS.432.2909F} {432, 2909}

\bibitem[\protect\citeauthoryear{{George} et~al.,}{{George}
  et~al.}{2015}]{2015ApJ...799..177G}
{George} E.~M.,  et~al., 2015, \mn@doi [\apj] {10.1088/0004-637X/799/2/177},
  \href {http://adsabs.harvard.edu/abs/2015ApJ...799..177G} {799, 177}

\bibitem[\protect\citeauthoryear{{Greig} \& {Mesinger}}{{Greig} \&
  {Mesinger}}{2016}]{2016arXiv160505374G}
{Greig} B.,  {Mesinger} A.,  2016, preprint, \href
  {http://adsabs.harvard.edu/abs/2016arXiv160505374G} {} (\mn@eprint {arXiv}
  {1605.05374})

\bibitem[\protect\citeauthoryear{{Haiman} \& {Bryan}}{{Haiman} \&
  {Bryan}}{2006}]{Haiman06}
{Haiman} Z.,  {Bryan} G.~L.,  2006, \mn@doi [\apj] {10.1086/506580}, \href
  {http://adsabs.harvard.edu/abs/2006ApJ...650....7H} {650, 7}

\bibitem[\protect\citeauthoryear{{Haiman} \& {Holder}}{{Haiman} \&
  {Holder}}{2003}]{2003ApJ...595....1H}
{Haiman} Z.,  {Holder} G.~P.,  2003, \mn@doi [\apj] {10.1086/377337}, \href
  {http://adsabs.harvard.edu/abs/2003ApJ...595....1H} {595, 1}

\bibitem[\protect\citeauthoryear{{Haiman}, {Rees}  \& {Loeb}}{{Haiman}
  et~al.}{1997}]{1997ApJ...476..458H}
{Haiman} Z.,  {Rees} M.~J.,   {Loeb} A.,  1997, \apj, \href
  {http://adsabs.harvard.edu/abs/1997ApJ...476..458H} {476, 458}

\bibitem[\protect\citeauthoryear{{Heinrich}, {Miranda}  \& {Hu}}{{Heinrich}
  et~al.}{2016}]{2016arXiv160904788H}
{Heinrich} C.~H.,  {Miranda} V.,   {Hu} W.,  2016, preprint, \href
  {http://adsabs.harvard.edu/abs/2016arXiv160904788H} {} (\mn@eprint {arXiv}
  {1609.04788})

\bibitem[\protect\citeauthoryear{{Hinshaw} et~al.,}{{Hinshaw}
  et~al.}{2013}]{2013ApJS..208...19H}
{Hinshaw} G.,  et~al., 2013, \mn@doi [\apjs] {10.1088/0067-0049/208/2/19},
  \href {http://adsabs.harvard.edu/abs/2013ApJS..208...19H} {208, 19}

\bibitem[\protect\citeauthoryear{{Hogan}, {Kaiser}  \& {Rees}}{{Hogan}
  et~al.}{1982}]{1982RSPTA.307...97H}
{Hogan} C.~J.,  {Kaiser} N.,   {Rees} M.~J.,  1982, \mn@doi [Philosophical
  Transactions of the Royal Society of London Series A]
  {10.1098/rsta.1982.0104}, \href
  {http://adsabs.harvard.edu/abs/1982RSPTA.307...97H} {307, 97}

\bibitem[\protect\citeauthoryear{Hu \& Holder}{Hu \& Holder}{2003}]{Hu:2003gh}
Hu W.,  Holder G.~P.,  2003, \mn@doi [Phys. Rev.] {10.1103/PhysRevD.68.023001},
  D68, 023001

\bibitem[\protect\citeauthoryear{Hui \& Gnedin}{Hui \&
  Gnedin}{1997}]{Hui:1997dp}
Hui L.,  Gnedin N.~Y.,  1997, Mon.Not.Roy.Astron.Soc., 292, 27

\bibitem[\protect\citeauthoryear{{Kaurov}, {Hooper}  \& {Gnedin}}{{Kaurov}
  et~al.}{2015}]{2015arXiv151200526K}
{Kaurov} A.~A.,  {Hooper} D.,   {Gnedin} N.~Y.,  2015, preprint, \href
  {http://adsabs.harvard.edu/abs/2015arXiv151200526K} {} (\mn@eprint {arXiv}
  {1512.00526})

\bibitem[\protect\citeauthoryear{{Koh} \& {Wise}}{{Koh} \&
  {Wise}}{2016}]{2016arXiv160904400K}
{Koh} D.,  {Wise} J.~H.,  2016, preprint, \href
  {http://adsabs.harvard.edu/abs/2016arXiv160904400K} {} (\mn@eprint {arXiv}
  {1609.04400})

\bibitem[\protect\citeauthoryear{{Lidz}}{{Lidz}}{2016}]{Lidz16}
{Lidz} A.,  2016, in {Mesinger} A.,  ed.,  Astrophysics and Space Science
  Library Vol. 423, Understanding the Epoch of Cosmic Reionization: Challenges
  and Progress. p.~23 (\mn@eprint {arXiv} {1511.01188}),
  \mn@doi{10.1007/978-3-319-21957-8_2}

\bibitem[\protect\citeauthoryear{{Liu}, {Pritchard}, {Allison}, {Parsons},
  {Seljak}  \& {Sherwin}}{{Liu} et~al.}{2016}]{2016PhRvD..93d3013L}
{Liu} A.,  {Pritchard} J.~R.,  {Allison} R.,  {Parsons} A.~R.,  {Seljak} U.,
  {Sherwin} B.~D.,  2016, \mn@doi [\prd] {10.1103/PhysRevD.93.043013}, \href
  {http://adsabs.harvard.edu/abs/2016PhRvD..93d3013L} {93, 043013}

\bibitem[\protect\citeauthoryear{{Loeb} \& {Furlanetto}}{{Loeb} \&
  {Furlanetto}}{2013}]{Loeb13}
{Loeb} A.,  {Furlanetto} S.~R.,  2013, {The First Galaxies in the Universe}

\bibitem[\protect\citeauthoryear{{Madau}, {Haardt}  \& {Rees}}{{Madau}
  et~al.}{1999}]{Madau99}
{Madau} P.,  {Haardt} F.,   {Rees} M.~J.,  1999, \mn@doi [\apj]
  {10.1086/306975}, \href {http://adsabs.harvard.edu/abs/1999ApJ...514..648M}
  {514, 648}

\bibitem[\protect\citeauthoryear{{McQuinn}, {Oh}  \&
  {Faucher-Gigu{\`e}re}}{{McQuinn} et~al.}{2011}]{McQuinn:2011aa}
{McQuinn} M.,  {Oh} S.~P.,   {Faucher-Gigu{\`e}re} C.-A.,  2011, \mn@doi [\apj]
  {10.1088/0004-637X/743/1/82}, \href
  {http://adsabs.harvard.edu/abs/2011ApJ...743...82M} {743, 82}

\bibitem[\protect\citeauthoryear{{Mitra}, {Choudhury}  \& {Ferrara}}{{Mitra}
  et~al.}{2015}]{2015MNRAS.454L..76M}
{Mitra} S.,  {Choudhury} T.~R.,   {Ferrara} A.,  2015, \mn@doi [\mnras]
  {10.1093/mnrasl/slv134}, \href
  {http://adsabs.harvard.edu/abs/2015MNRAS.454L..76M} {454, L76}

\bibitem[\protect\citeauthoryear{Mortonson \& Hu}{Mortonson \&
  Hu}{2008a}]{Mortonson:2007hq}
Mortonson M.~J.,  Hu W.,  2008a, \mn@doi [Astrophys. J.] {10.1086/523958}, 672,
  737

\bibitem[\protect\citeauthoryear{Mortonson \& Hu}{Mortonson \&
  Hu}{2008b}]{Mortonson:2008rx}
Mortonson M.~J.,  Hu W.,  2008b, \mn@doi [Astrophys. J.] {10.1086/593031}, 686,
  L53

\bibitem[\protect\citeauthoryear{{Park}, {Shapiro}, {Komatsu}, {Iliev}, {Ahn}
  \& {Mellema}}{{Park} et~al.}{2013}]{2013ApJ...769...93P}
{Park} H.,  {Shapiro} P.~R.,  {Komatsu} E.,  {Iliev} I.~T.,  {Ahn} K.,
  {Mellema} G.,  2013, \mn@doi [\apj] {10.1088/0004-637X/769/2/93}, \href
  {http://adsabs.harvard.edu/abs/2013ApJ...769...93P} {769, 93}

\bibitem[\protect\citeauthoryear{{Pawlik}, {Schaye}  \& {van
  Scherpenzeel}}{{Pawlik} et~al.}{2009}]{Pawlik:2008mr}
{Pawlik} A.~H.,  {Schaye} J.,   {van Scherpenzeel} E.,  2009, \mn@doi [\mnras]
  {10.1111/j.1365-2966.2009.14486.x}, \href
  {http://adsabs.harvard.edu/abs/2009MNRAS.394.1812P} {394, 1812}

\bibitem[\protect\citeauthoryear{{Planck Collaboration} et~al.,}{{Planck
  Collaboration} et~al.}{2016a}]{2016arXiv160503507P}
{Planck Collaboration} et~al., 2016a, preprint, \href
  {http://adsabs.harvard.edu/abs/2016arXiv160503507P} {} (\mn@eprint {arXiv}
  {1605.03507})

\bibitem[\protect\citeauthoryear{{Planck Collaboration} et~al.,}{{Planck
  Collaboration} et~al.}{2016b}]{2015arXiv150201589P}
{Planck Collaboration} et~al., 2016b, \mn@doi [\aap]
  {10.1051/0004-6361/201525830}, \href
  {http://adsabs.harvard.edu/abs/2016A%26A...594A..13P} {594, A13}

\bibitem[\protect\citeauthoryear{{Ricotti} \& {Ostriker}}{{Ricotti} \&
  {Ostriker}}{2004}]{2004MNRAS.352..547R}
{Ricotti} M.,  {Ostriker} J.~P.,  2004, \mn@doi [\mnras]
  {10.1111/j.1365-2966.2004.07942.x}, \href
  {http://adsabs.harvard.edu/abs/2004MNRAS.352..547R} {352, 547}

\bibitem[\protect\citeauthoryear{{Robertson}, {Ellis}, {Furlanetto}  \&
  {Dunlop}}{{Robertson} et~al.}{2015}]{2015ApJ...802L..19R}
{Robertson} B.~E.,  {Ellis} R.~S.,  {Furlanetto} S.~R.,   {Dunlop} J.~S.,
  2015, \mn@doi [\apjl] {10.1088/2041-8205/802/2/L19}, \href
  {http://adsabs.harvard.edu/abs/2015ApJ...802L..19R} {802, L19}

\bibitem[\protect\citeauthoryear{{Schaerer}}{{Schaerer}}{2002}]{Schaerer02}
{Schaerer} D.,  2002, \mn@doi [\aap] {10.1051/0004-6361:20011619}, \href
  {http://adsabs.harvard.edu/abs/2002A%26A...382...28S} {382, 28}

\bibitem[\protect\citeauthoryear{{Shapiro} \& {Giroux}}{{Shapiro} \&
  {Giroux}}{1987}]{Shapiro87}
{Shapiro} P.~R.,  {Giroux} M.~L.,  1987, \mn@doi [\apjl] {10.1086/185015},
  \href {http://adsabs.harvard.edu/abs/1987ApJ...321L.107S} {321, L107}

\bibitem[\protect\citeauthoryear{Sheth, Mo  \& Tormen}{Sheth
  et~al.}{2001}]{Sheth:1999su}
Sheth R.~K.,  Mo H.,   Tormen G.,  2001, \mn@doi [Mon.Not.Roy.Astron.Soc.]
  {10.1046/j.1365-8711.2001.04006.x}, 323, 1

\bibitem[\protect\citeauthoryear{{Sun} \& {Furlanetto}}{{Sun} \&
  {Furlanetto}}{2016}]{2016MNRAS.460..417S}
{Sun} G.,  {Furlanetto} S.~R.,  2016, \mn@doi [\mnras] {10.1093/mnras/stw980},
  \href {http://adsabs.harvard.edu/abs/2016MNRAS.460..417S} {460, 417}

\bibitem[\protect\citeauthoryear{{The COrE Collaboration} et~al.,}{{The COrE
  Collaboration} et~al.}{2011}]{2011arXiv1102.2181T}
{The COrE Collaboration} et~al., 2011, preprint, \href
  {http://adsabs.harvard.edu/abs/2011arXiv1102.2181T} {} (\mn@eprint {arXiv}
  {1102.2181})

\bibitem[\protect\citeauthoryear{{Tseliakhovich} \& {Hirata}}{{Tseliakhovich}
  \& {Hirata}}{2010}]{2010PhRvD..82h3520T}
{Tseliakhovich} D.,  {Hirata} C.,  2010, \mn@doi [\prd]
  {10.1103/PhysRevD.82.083520}, \href
  {http://adsabs.harvard.edu/abs/2010PhRvD..82h3520T} {82, 083520}

\bibitem[\protect\citeauthoryear{{Tumlinson} \& {Shull}}{{Tumlinson} \&
  {Shull}}{2000}]{2000ApJ...528L..65T}
{Tumlinson} J.,  {Shull} J.~M.,  2000, \mn@doi [\apjl] {10.1086/312432}, \href
  {http://adsabs.harvard.edu/abs/2000ApJ...528L..65T} {528, L65}

\bibitem[\protect\citeauthoryear{{Visbal}, {Haiman}, {Terrazas}, {Bryan}  \&
  {Barkana}}{{Visbal} et~al.}{2014}]{2014MNRAS.445..107V}
{Visbal} E.,  {Haiman} Z.,  {Terrazas} B.,  {Bryan} G.~L.,   {Barkana} R.,
  2014, \mn@doi [\mnras] {10.1093/mnras/stu1710}, \href
  {http://adsabs.harvard.edu/abs/2014MNRAS.445..107V} {445, 107}

\bibitem[\protect\citeauthoryear{{Visbal}, {Haiman}  \& {Bryan}}{{Visbal}
  et~al.}{2015}]{Visbal15}
{Visbal} E.,  {Haiman} Z.,   {Bryan} G.~L.,  2015, \mn@doi [\mnras]
  {10.1093/mnras/stv1941}, \href
  {http://adsabs.harvard.edu/abs/2015MNRAS.453.4456V} {453, 4456}

\bibitem[\protect\citeauthoryear{{Yoshida}, {Abel}, {Hernquist}  \&
  {Sugiyama}}{{Yoshida} et~al.}{2003}]{2003ApJ...592..645Y}
{Yoshida} N.,  {Abel} T.,  {Hernquist} L.,   {Sugiyama} N.,  2003, \mn@doi
  [\apj] {10.1086/375810}, \href
  {http://adsabs.harvard.edu/abs/2003ApJ...592..645Y} {592, 645}

\bibitem[\protect\citeauthoryear{{Zahn} et~al.,}{{Zahn}
  et~al.}{2012}]{2012ApJ...756...65Z}
{Zahn} O.,  et~al., 2012, \mn@doi [\apj] {10.1088/0004-637X/756/1/65}, \href
  {http://adsabs.harvard.edu/abs/2012ApJ...756...65Z} {756, 65}

\bibitem[\protect\citeauthoryear{{Zaldarriaga}}{{Zaldarriaga}}{1997}]{1997PhRvD..55.1822Z}
{Zaldarriaga} M.,  1997, \mn@doi [\prd] {10.1103/PhysRevD.55.1822}, \href
  {http://adsabs.harvard.edu/abs/1997PhRvD..55.1822Z} {55, 1822}

\makeatother
\end{thebibliography}
\bsp	
\label{lastpage}
\end{document}